\documentclass[aps, prd, reprint, showpacs, superscriptaddress]{revtex4-1}
\usepackage{float}
\usepackage{microtype}
\usepackage{amsmath,amssymb}
\usepackage{linenoAMS}
\usepackage[usenames, dvipsnames, svgnames, table]{xcolor}
\usepackage[colorlinks=true, citecolor=RoyalBlue,
            linkcolor=BrickRed, urlcolor=ForestGreen]{hyperref}
\usepackage{graphicx}
\usepackage{siunitx}
\usepackage[capitalise]{cleveref}

\usepackage{linenoAMS}
\usepackage{amsmath}
\usepackage{comment}
\usepackage[
    range-phrase=--,
    range-units=single,
    retain-explicit-plus=true,
    retain-unity-mantissa=false,
]{siunitx}

\usepackage[all]{hypcap}
\usepackage{enumitem}
\usepackage{braket}
\usepackage{comment}
\usepackage{mathtools}

\graphicspath{{figures/}}
\usepackage[capitalise]{cleveref}
\usepackage{graphicx}
\usepackage{xfrac}
\usepackage{ulem}
\usepackage{cancel}

\definecolor{spring}{rgb}{0.7,0.9,0.7}
\definecolor{brick}{rgb}{0.7,0.2,0.1}
\definecolor{redHL}{rgb}{1.0,0.7,0.7}
\definecolor{blueHL}{rgb}{0.7,0.7,1.0}
\definecolor{ElectricBlue}{rgb}{0.49, 0.976, 1.0}
\definecolor{blueT}{rgb}{0.039, 0.729, 0.71}
\definecolor{DarkGreen}{rgb}{0, 0, 0}
\definecolor{blue}{rgb}{0.066, 0.310, 0.984}
\definecolor{blueT}{rgb}{0.039, 0.729, 0.71}
\definecolor{black}{rgb}{0, 0, 0}
\definecolor{violet}{rgb}{0.749, 0.467, 0.965}

\usepackage[capitalise]{cleveref}
\usepackage{commands}
\begin{document}
\title{Direct measurement of the \tripletPzero~clock state natural lifetime in \Sra~}
\author{Jonathan Dolde}%
\thanks{These authors contributed equally}
\author{Dhruva Ganapathy}%
\thanks{These authors contributed equally}
\author{Xin Zheng}
\author{Shuo Ma}%
\affiliation{Department of Physics, University of California, Berkeley, CA 94720, USA
}%

\author{Kyle Beloy}
\affiliation{National Institute of Standards and Technology, 325 Broadway, Boulder, Colorado 80305, USA}

\author{Shimon Kolkowitz}
\email{To whom correspondence should be addressed; \\
Email: kolkowitz@berkeley.edu}
\affiliation{Department of Physics, University of California, Berkeley, CA 94720, USA
}%

\overfullrule 0pt 
\parskip0pt
\hyphenpenalty9999

\date{\today}
\begin{abstract}

Optical lattice clocks based on the narrow (5s$^{2}$)\singletS-(5s5p)\tripletPzero transition in neutral strontium (Sr) are among the most precise and accurate measurement devices in existence. Although this transition is completely forbidden by selection rules, state mixing from the hyperfine interaction in \Sra~provides a weakly allowed transition that can be coherently driven with practical clock laser intensities. While the coherent interrogation times of optical clocks are typically set by the linewidth of the probe laser, this limitation can be overcome in synchronous differential comparisons between ensembles. In such measurements the natural lifetime of the \singletS-\tripletPzero clock transition becomes the fundamental limiting factor to the duration of a single run of the experiment. However, a direct measurement of the decay rate of the clock excited state is quite challenging due to the competing effects of other loss channels such as Raman scattering, inelastic collisions and atom-loss due to background gas. In this work, we monitor the decay of Sr atoms trapped in an optical lattice and initialized in the \tripletPzero state. By making measurements of high and low density ensembles of both \Sra~and \Srb~across varying lattice trap depths, we isolate radiative decay, which accounts for a significant fraction of the observed decays at low depths. We obtain a natural radiative decay lifetime of \lifetime~s  for the \tripletPzero clock state in \Sra, a value that is consistent with previously reported measurements and theoretical predictions. We also introduce an additional measurement scheme that involves repeated measurements of the ground state population within a single experimental sequence, validating our model and the consistency of the measured rates. We expect that the techniques introduced in this work are applicable to lifetime measurements of long-lived metastable states in other atoms and ions used for clocks and quantum computing. The results presented in this work inform performance estimates of proposed gravitational wave detectors and long-baseline atom interferometers making use of \Sra. 
\end{abstract}

\maketitle



\section{\label{sec:intro}Introduction}
 
Over the past two decades optical atomic clocks have rapidly developed into the most precise and accurate frequency references ever realized \cite{Ludlow15}, with clocks based on neutral strontium (Sr) now demonstrating fractional frequency accuracy at the level of \recordperf~\cite{Aeppli24}, and clock instabilities down to $4.8\times10^{-17}/\sqrt{\textrm{Hz}}$ \cite{oelker2019demonstration}. These Sr optical lattice clocks make use of the transition between the 5s$^{2}$ \singletS and 5s5p \tripletPzero state, both of which are magnetically insensitive states due to their lack of electronic angular momentum. While this transition is completely forbidden by selection rules, hyperfine interactions from the nuclear spin of \Sra~induce state mixing within the 5s5p \tripletPj manifold, creating a narrow optical transition near $429$~THz~\cite{Courtillot03} that balances the competing requirements of a long excited state lifetime with a sufficient transition dipole moment to enable coherent excitation with moderate clock laser intensities. \Sra~based systems have therefore now been used for tests of the gravitational redshift down to cm and mm scales \cite{Takamoto20, Bothwell22, Zheng23}, to constrain ultra-light dark matter coupling \cite{Kennedy20, Beloy21}, for clock frequency ratio measurements toward a redefinition of the SI second \cite{Beloy21, Dorscher21, Hisai21}, and for relativistic geodesy \cite{Takano2016, Grotti18, Grotti24}, and have also been proposed as leading candidates for long-baseline optical clock atom interferometers \cite{Abe21,abend2024terrestrial} and quantum computing \cite{barnes2022assembly}.

The ultimate limit to the achievable coherent interrogation time of a clock measurement is set by the lifetime of the excited clock state. Theory calculations predict a lifetime for the \tripletPzero state in \Sra~of \textgreater 100 s, with calculated values ranging from \theolower~s to \theoupper~s \cite{Santra04, Lu23}. Experimentally measuring this value more precisely will allow for better estimations of the sensitivities and bandwidths that can be achieved in existing and future platforms that make use of the \Sra~clock transition. For example, the excited state lifetime is a contributing factor in estimating the sensitivities of future clock-based gravitational wave detectors in space \cite{Kolkowitz16}. 

An accurate measurement of such a long excited state lifetime is made challenging by competing decay processes in an optical trap. These include optical lattice trap-induced Raman scattering, black body radiation (BBR) induced scattering, background gas collisions, inelastic scattering from collisions between excited state atoms, and other mechanisms of loss of atoms from the lattice. To the best of our knowledge, the only previously reported direct measurement was made by \dors et al.~\cite{Dorscher18}, in which \Sra~atoms were loaded into a 1-D optical lattice trap, initialized in the excited state, and held for varying time intervals, after which the populations of the ground and excited states were measured. However, the extraction of the \tripletPzero lifetime from these measurements was particularly constrained by the vacuum-limited lifetime of \dorvaclife~s, which was far below their reported value of \dorlife~s.

While the \Sra~lifetime can be inferred from a measurement of the dipole matrix element of the \singletS to \tripletPzero transition, the two other reported values \cite{Muniz21, Lu24} that use this method disagree with each other well beyond what would be expected based on their statistical and systematic uncertainties, indicating that one or both of these measurements may be limited by additional systematic uncertainties, and motivating direct measurements of the quantity of interest.

In this work, we perform direct lifetime measurements similar to those of \dors et al. However, we take advantage of significantly longer trapped-atom lifetimes (\glifelow$-$\glifehigh~s), and also introduce the concept of making simultaneous measurements of multiple ensembles at high and low density, and with both \Sra~and \Srb~. This allows us to isolate the contribution of radiative decay from other competing rates, obtaining a natural lifetime of \lifetime~s for the \tripletPzero state in \Sra. Inspired by the concepts of ``erasure conversion" and mid-circuit measurement from quantum computing with alkaline-earth atoms, we also confirm the consistency of our measurements and the extracted rates by performing novel measurements in which the population decaying into the ground state is repeatedly measured while the population remaining in the excited state is left undisturbed and continues to decay. 

\begin{figure*}
    \centering
    \includegraphics[width=\linewidth]{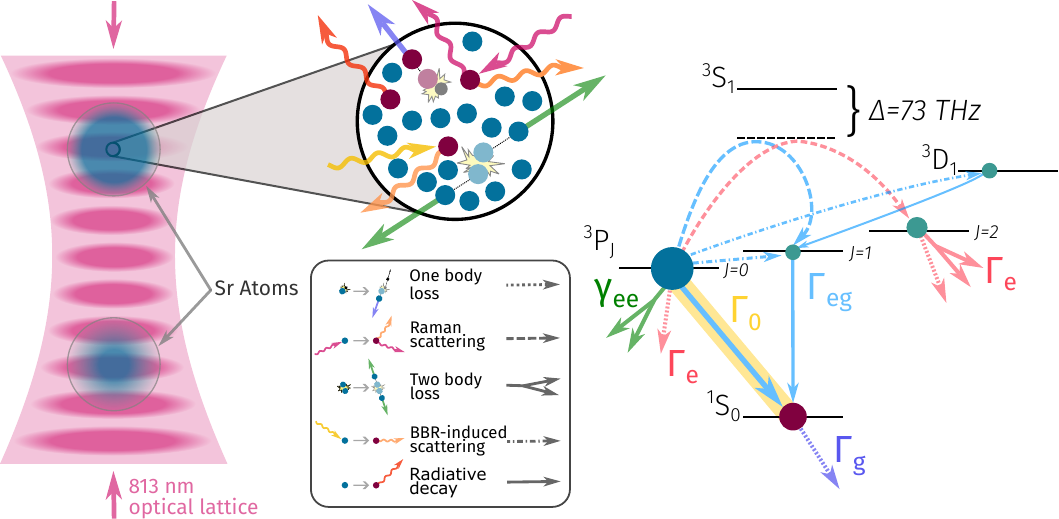}
    \caption{Population decay in an optical lattice. Two ensembles of \Sra~or \Srb~atoms are loaded into an optical lattice (left) and prepared in the excited \tripletPzero state. The zoomed-in portion of the atomic ensemble shows atoms that have decayed into the \singletS ground state (red) and atoms that remain in the excited state (blue), along with the decay and loss mechanisms considered in our model. These processes are also represented using different line styles in a Sr level diagram (right). The line colors correspond to different parameters in the rate equation (see \cref{eq:rate_equations}). These include one-body losses, $\gloss$ and $\eloss$,  from the ground and excited states, two-body loss, \twobodyloss, and decay rate from excited to ground state, $\decayrate$. The radiative decay rate, $\radiativedecay$, corresponding to the natural lifetime of the clock state, is highlighted in yellow.}
    \label{fig:figure_1}
\end{figure*}

\begin{figure*}
{\includegraphics[width=1.0\linewidth]{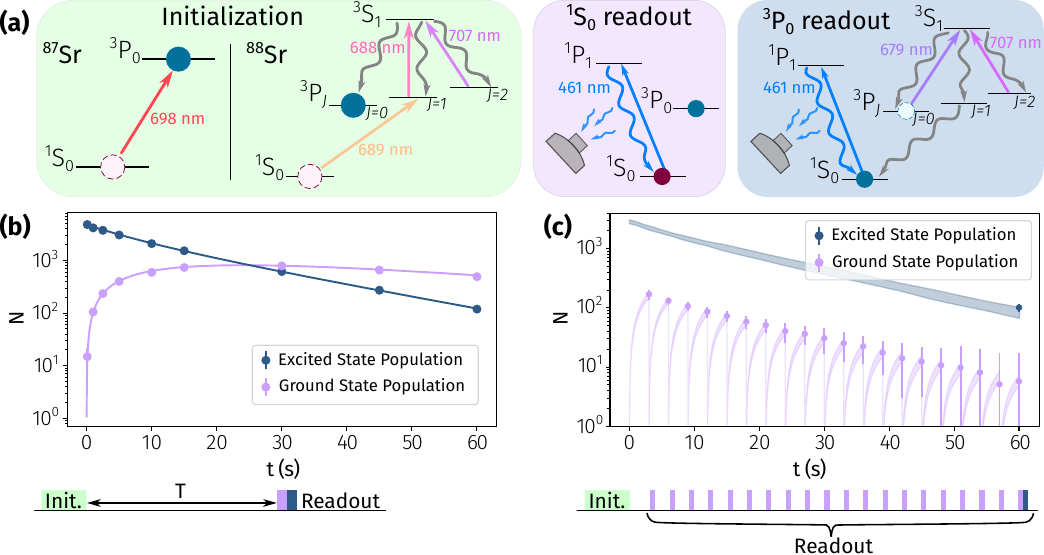}}

\caption{Population decay measurement scheme and example data for \Sra~loaded at 40 \erec. \textbf{a)} After loading, \Sra~atoms are directly initialized in the \tripletPzero excited clock state using a $\pi$-pulse. In \Srb, the atoms are first initialized in the \tripletPone~state, then are optically pumped into \tripletPzero via the \tripletS~state. The readout procedure is the same for \Sra~and \Srb; the \singletS ground state population is read out using fluorescence from the \singletP state after a 461 nm cycling pulse. The excited state population is excited to the \tripletS state where it decays to the ground state via the \tripletPone state and is then read out. \textbf{b)} In the standard measurement, the atoms are prepared in the excited state using a $\pi$-pulse and allowed to decay for time T, after which the ground (purple) and excited (blue) state populations are read out destructively. This process is repeated while varying T, and the resulting decay curves are fit to the rate equations for the state populations. The data points and error bars result from averaging over 10 sets of population data. This representative plot is of high-density \Sra~atoms held at a trap depth of $\approx$ 40\erec. \textbf{c)} In the multi-readout measurement, the ground state population is read out at regular time intervals within a single experiment sequence, while atoms in the excited state are unaffected and continue to decay. The data points and error bars are obtained by median averaging the data over 48 experimental runs. The final readout destructively measures populations of both states. In order to estimate the initial population of the excited state, the atoms are prepared using a $\pi/2$ pulse and the remaining ground state population is destructively read out immediately at $t=0$. The model curves use parameters inferred from the standard measurement. The shaded region indicates the uncertainty in model parameters.} 
\label{fig:data}
\end{figure*}

\section{\label{sec:exp}Experimental Setup}
In order to measure the lifetime of the \tripletPzero~state of \Sra, we load ensembles of \Sra~or \Srb~into a vertically oriented 1-D optical lattice at the `magic' wavelength of 813.4 nm (see \cref{fig:figure_1}) for the Sr clock transition \cite{Zheng22}. Our loading sequence starts with a standard two-stage magneto-optical trap (MOT), which cools the atoms to $\sim$1 \uK. As the atoms are cooled, they fall into the overlapping optical lattice potential. Multiple spatially resolved atomic ensembles are then loaded into different regions of the lattice by modulating the zero-field point of the narrow-line second-stage MOT with additional bias coils \cite{Niroula24}. The atom number in each ensemble is controlled by varying the duration of time for which the narrow-line MOT is overlapped with the lattice at each location. We operate with two ensembles separated by 0.8 millimeters, and with a factor of three difference in atom numbers in order to separate out the effects of density-dependent two-body loss. Once these ensembles are loaded into the lattice, we employ sideband and radial cooling and then ramp the lattice intensity from the loading depth of $\sim$70 \erec~to $\sim$8 \erec~to filter out hotter atoms. The recoil energy is given by \erec = $h^{2}$/(2m$_{87}$$\lambda^{2}$), where $m_{87}$ is the atomic mass of \Sra, $\lambda = 813$ nm is the wavelength of the lattice laser and $h$ is Planck's constant.  The lattice is then ramped back up to our operational depth, between $\sim$(20 and 65)\erec. Finally, we initialize our ensembles into the excited clock state, $\ket{^3\textrm{P}_0,~m_F=3/2}$, as shown in \cref{fig:data}a. For \Sra, we initialize into the excited state with a coherent $\pi-$pulse on the \singletS to \tripletPzero transition. In \Srb, this transition is completely forbidden in the absence of a magnetic field, and the dipole moment remains too weak to drive at the magnetic fields we currently achieve in our apparatus. We instead drive the \singletS$-$\tripletPone, \tripletPone$-$\tripletS, and \tripletPtwo$-$\tripletS transitions simultaneously in order to optically pump the atoms into the excited state. After initialization into the excited state we use a strong 461 nm pulse on the \singletS$-$\singletP transition to clear any residual ground state atoms out of the trap. To extract the radial temperature of the atoms, we direct a probe beam radially at ground state atoms and measure their Doppler-broadened excitation profile. For \Sra, we probe using  the \singletS$-$\tripletPzero~clock transition, but in \Srb, we cannot directly use this transition, so we use \singletS$-$\tripletPone. To account for heating during loading of \Srb, we measure the radial temperature after performing one cycle of loading in and out of \tripletPzero~as described above.

To measure the decay curves shown in \cref{fig:data}b, we hold atoms in the lattice for a variable time T, after which the populations of the ground and excited states are destructively read out in series with florescence imaging using a 461 nm probe beam.  By performing this experiment across different lattice trap depths between $\sim$(20 and 65)~\erec, we observe how the decay rate from the excited to ground state is affected by Raman scattering induced by the 813 nm trapping light. Simultaneous loading of two ensembles with a factor of 3 difference in atom number is used to check for consistency between different initial densities during the same experiment.

We also perform a modified experiment that is meant to reduce the effects of run-to-run differences in initial atom number and to confirm the consistency of the extracted rates. We initialize atoms into the excited state, but then repeatedly image the atoms that have decayed or scattered into the ground state during the lattice hold time while leaving the excited state undisturbed. This imaging is repeated with a constant interval, $t_{\text{rep}}$, between 3 and 5 seconds depending on trap depth. We thereby  measure almost all atoms that decay to the ground state, as these atoms are quickly counted before they have a chance to escape the lattice. This imaging protocol leaves population of the excited state intact, such that we construct the entire ground state population curve with a single initialization of excited state atoms. In order to determine the initial population of the excited state ($t=0$), we reduce the duration of the initialization pulse such that a specifically calibrated ratio of atoms are prepared in the excited state, with the rest remaining in the ground state. By then immediately imaging the leftover ground state atoms, we estimate the number of atoms in the excited state. We then only directly read out the population of the excited state at the end of the sequence. We refer to this second experimental method as the ``multi-readout" technique to differentiate from the previously described ``standard" or ``single-imaging" technique. Since this method captures multiple snapshots of the ground state population throughout a single experiment, we construct a ground state population curve (see \cref{fig:data}c) with only a single experimental run, which both speeds up data taking and reduces the effect of run-to-run variance in initial atom number.

\begin{figure}
    \centering
    \includegraphics[width=\linewidth]{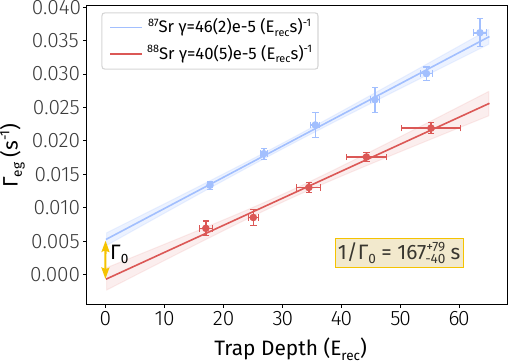}
    \caption{Decay rate $\decayrate$ from the excited \tripletPzero state to the ground \singletS state in \Sra~(blue) and \Srb~(red) as a function of lattice trap depth. The data are fit to \cref{eq:decay_linear} to extract the Raman scattering rate from the excited to ground state. The natural lifetime (1/\radiativedecay) of \Sra~is given by the inverse of the difference between the decay rates of the two isotopes extrapolated to 0 \erec. The error bars on the lifetime are obtained by fitting resampled data to obtain a distribution of model parameters. The x-axis is an effective depth, where the radial profile of the trapping light is factored into the total intensity the atoms experience. The x-axis errors in this measurement arise mostly from the uncertainty in the determination of the radial temperature of the atomic ensembles.}
    \label{fig:lifetime}
\end{figure}
\section{\label{sec:results}Model and Results}
To account for the various sources of decay and loss in the optical lattice, we use the model diagrammatically represented in \cref{fig:figure_1}. We consider the following rate equations: 
\begin{equation}
\begin{aligned}
\dot{\Ne} &= -(\eloss+\decayrate)N_e-\twobodyloss N_e^2, \\ \dot{\Ng} &= -\gloss \Ng+\decayrate\Ne,
\label{eq:rate_equations}
\end{aligned}
\end{equation}
where the $N_g$ is the ground state population, $N_e$ is the excited clock state population, $\decayrate$ is the rate of decay from the excited to ground state, $\gloss$ is the background loss rate from the ground state, $\eloss$ is the background loss rate from the excited state, $\twobodyloss N_{e} = K_{ee}/V$ is the density dependent two-body loss rate, where $K_{ee}$ is the e-e collision rate and V is the effective trapping volume (see \cref{app:two_body}), and the dot signifies a derivative with respect to time. The decay rate can further be expanded as: 
\begin{equation}
    \decayrate = (\radiativedecay+\gamma U),
    \label{eq:decay_linear}
\end{equation}
where $\radiativedecay$ is the radiative decay rate, $\gamma$ is the Raman scattering rate per unit trap depth and $U$ is the effective optical trap depth, calculated by using the measured axial trap depth, $U_{\text{axial}}$, and radial atomic temperature, $T_{r}$, through \cite{Siegel24}
\begin{equation}
    U = U_{\text{axial}}(1 + k_{B}T_{r}/U_{\text{axial}})^{-1}.
\end{equation}

We fit our model to standard readout data, averaged over 10 sets of population measurements (\cref{fig:data}b) for both Sr isotopes \cref{eq:rate_equations}. While using the multi-readout data (\cref{fig:data}c) to infer model parameters is challenging due to the lack of $N_e$ data, we instead use it to verify the consistency of the model and rates extracted from the other measurements. To numerically solve the equations for the multi-readout technique, $N_g$ is set to zero after every readout while $N_e$ remains unchanged (here, we use the data corresponding to the higher density ensemble; we discuss the low-density data in \cref{app:low_density}). Additional fit parameters are discussed in \cref{app:params}. The uncertainty in the fit parameters is estimated by resampling the data according to the distribution of population measurements. 

We then plot the extracted values of $\decayrate$ as a function of depth and a linear fit allows us to extract the Raman scattering rate for \tripletPzero to \tripletPone as well as the decay rate extrapolated to a depth of 0 \erec. This should correspond to $\radiativedecay$, along with additional contributions from black-body scattering that is common to both \Sra~and \Srb~. As radiative decay is expected to be absent in \Srb, we subtract the inferred value of $\decayrate$ for \Srb~from that of \Sra~in order to obtain  $\radiativedecay$ for \Sra. These results are plotted in \cref{fig:lifetime}. By simulating experimental data based on the model we use to fit to observed data, we verify that the systematic uncertainty due to covariance in fit parameters is negligible compared to the statistical uncertainty, which is dominated by the run-to-run variation in the number of initially loaded atoms. The other major systematic contribution to the uncertainty in the linear fits in \cref{fig:lifetime} arises from the effective trap depth uncertainty, which is ultimately limited by our measurements of the atomic temperature.

\begin{figure}
    \centering
    \includegraphics[width=1.0\linewidth]{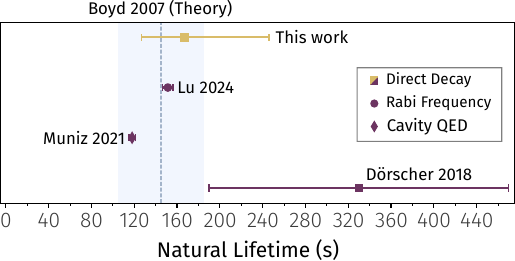}
    \caption{Comparison of measured \Sra~lifetime values from Refs.~Lu 2024 \cite{Lu24}, Muniz 2021\cite{Muniz21}, Dörscher 2018 \cite{Dorscher18}. This work and Dörscher both measure the \tripletPzero lifetime through observations of radiative decay. Lu and Muniz both measured the dipole moment of the clock transition, leading to their lifetime values. Muniz used a cavity QED comparison between \singletS to \tripletPzero and \singletS to \tripletPone dipole moments, and Lu measured the Rabi frequency of the transition in conjunction with characterization of the laser intensity. The dashed line and shaded region indicate theoretical lifetime value and uncertainty, respectively, calculated in Boyd et al \cite{Boyd07}. Boyd applied Breit-Wills theory to calculate the lifetime from measured hyperfine splitting values. The theory uncertainty arises from the choice of theoretical framework and confidence associated with lifetimes of the \singletP and \tripletPone states.}
    \label{fig:comparison}
\end{figure}

We measure Raman scattering rates $\gamma$, of \ramana~and \ramanb~for \Sra~and \Srb~respectively, which are in agreement with each other and with previously reported values of $\sim$\ramanlit~ \cite{Dorscher18,Hutson19}. For \Srb~at 0 \erec, we find \decayrate~to be \interceptb, which is roughly $2\sigma$ from the theoretically predicted black-body-induced scattering rate at 300 K of \bbrrate (see \cref{app:bbrrate}). This discrepancy may be the result of statistical fluctuations, or could arise from an unaccounted for systematic effect. However, as we subtract the intercept of the \Srb~ rate from that of \Sra, any systematic that is common to both will cancel out in our extracted radiative decay rate. The calculated black-body-induced scattering rate also subtracts out and does not contribute. The resulting radiative decay rate we obtain for the excited clock state in \Sra~is $6.0(19)\times10^{-3}$ s$^{-1}$ corresponding to a lifetime of \lifetime~s. Note that our measurement and analysis method results in a radiative decay rate with symmetric uncertainties. Inverting the measured decay rate then results in asymmetric uncertainties in the extracted radiative lifetime. We compare this value with previous measurements in \cref{fig:comparison}. The contributions of atom number and temperature uncertainties to the uncertainty in the extracted radiative decay rate are listed in \cref{tab:uncertainty}. 

The larger uncertainty in the measured \Srb~radial temperature arises primarily due to the fact that the relatively broad \singletS$-$\tripletPone~transition is used for this measurement.

\begin{table}[b]
    \centering
     \caption{Contribution of statistical and systematic uncertainties to the total uncertainty in extracted radiative decay rates.}
    \begin{tabular}{cc}
    \hline
    \hline
         \textbf{Source} & \textbf{Uncertainty ($\times10^{-3}$s$^{-1}$)} \\
         \hline
        \hline
         \textbf{\Sra}& 0.9\\
         \hline
         Atom number & 0.9\\
         Atom temperature & 0.3\\
         
         \hline
         \hline
         \textbf{\Srb}& 1.7\\
         \hline
         Atom number & 1.3\\
         Atom temperature & 0.8\\
         \hline
         \hline
         \textbf{Total} & 1.9\\
         \end{tabular}
          
    \label{tab:uncertainty}
    \end{table}
    
\section{\label{sec:conclusion}Conclusions and Outlook}
We have presented a measurement of the natural lifetime of the excited \tripletPzero~clock state in \Sra~through the direct observation of radiative decay. We find the lifetime to be \lifetime~s . This value is largely consistent with previous measurements by Lu et al. \cite{Lu24}, Dörscher et al. \cite{Dorscher18}, and Muniz et al. \cite{Muniz21}. It also lies within the theory uncertainty given by Boyd et al. \cite{Boyd07}.

By varying the lattice depth, ensemble density, and measuring with two different isotopes, we disentangled the radiative decay rate from the other decay processes occurring in a 1-D optical lattice trap. At lower trap depths, ultra-high vacuum and long atom lifetimes in the optical lattice allowed us to attribute a significant fraction ($\sim50\%$) of the measured ground state population from the excited state to radiative decay. 

The primary limitations in our measurement were the statistical uncertainty due to variation in the initial number of loaded atoms between experimental runs and uncertainty in the radial temperature of \Srb, both of which could be mitigated through technical improvements to the experiment in future work. To mitigate the systematic uncertainty arising from uncertainty in additional fit parameters, adding a resonant laser that rapidly repumps out of the \tripletPone state would suppress decay events from \tripletPzero to \singletS through the \tripletPone state due to Raman and black-body scattering. Finally, varying the effective trapping volume could help in isolating the two-body loss rates from the rest of the measurement.

The lifetime of the \tripletPzero state sets an ultimate limit on the coherence of the narrow-line clock transition of \Sra. Our measurement of this value helps to determine the achievable performance of comparisons between \Sra~clocks in applications where differential measurement techniques bypass the limits imposed by the local oscillator \cite{Zheng22, Bothwell22}. The experimental techniques described in this work could also be used to make direct lifetime measurements of other similarly long-lived atomic transitions in atoms and ions.

\noindent \textbf{\textit{Authors' note:}} In a complementary work performed in parallel to our own, \cite{kim2025atomic}, K.~Kim \textit{et al.}~report a value of $174(28)$~s for the radiative lifetime of the \tripletPzero clock state in \Sra, which is  fully consistent with the measured value reported in this work.

\section*{Acknowledgments}

We thank Jeff Thompson, James Thompson, Jun Ye, Nils Huntemann, Andrew Ludlow, Andrew Jayich, Eugene Knyazev, David Hume, Harikesh Ranganath and Jacob Siegel for fruitful discussions and insightful comments on the manuscript. 
This work was supported by a Packard Fellowship for Science and Engineering, the Army Research Office through agreement number W911NF-21-1-0012, the Sloan Foundation, the Simons Foundation, the Gordon and Betty Moore Foundation under grant DOI 10.37807/gbmf12966, NASA under grant No.~80NSSC24K1561, the National Institute of Standards and Technology/Physical Measurement Laboratory, and the National Science Foundation under Grants No.~2143870 and 2326810.


\appendix

\section{\label{app:params}Model Parameters}
Along with the excited state decay, the population data are used to estimate the other parameters of the rate equation. These parameters are listed in \cref{tab:params}.
\subsection{\label{app:one_body}One-Body Loss}

\begin{figure}
    \centering
    \includegraphics[width=\linewidth]{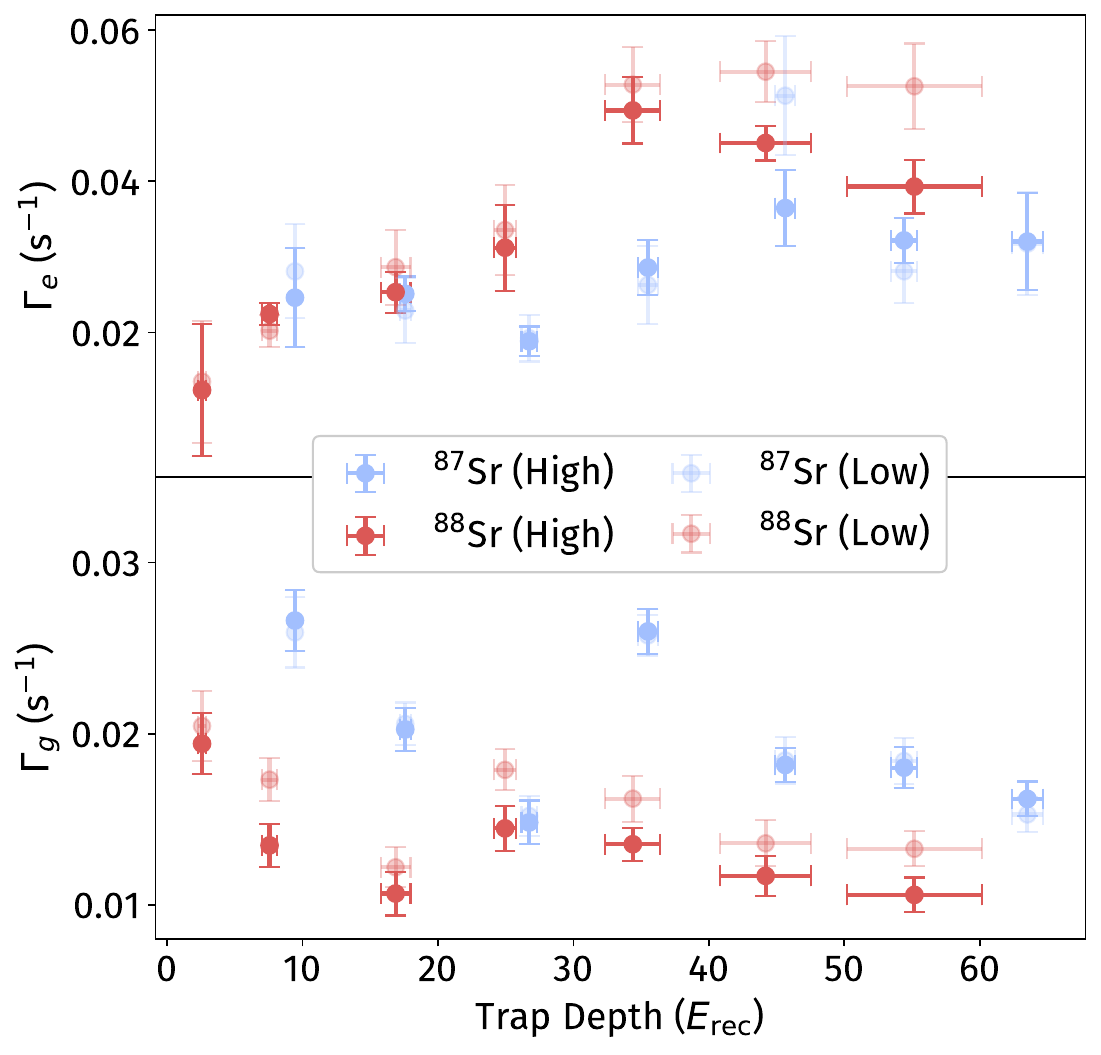}
    \caption{one-body loss rates for ground $\gloss$ and excited $\eloss$ states for high and low density ensembles of \Sra~and \Srb. $\gloss$ is measured independently while $\eloss$ is inferred by fitting standard readout data to the decay model (\cref{eq:rate_equations}).}
    \label{fig:loss_g}
\end{figure}

The one-body loss from the ground state, $\gloss$ is measured independently by loading atoms in the ground state and measuring the population decay. The extracted values and their uncertainties, shown in \cref{fig:loss_g}, are used to constrain fit parameters in the standard readout data analysis. The excited state loss is extracted from fitting the rate equations (\cref{eq:rate_equations}) to the standard readout data. 

\subsection{\label{app:two_body}Two-Body Loss}

In \cref{eq:rate_equations}, the two-body loss term, $\twobodyloss$, can be written as $K_{ee}/V$ where $V$ is the effective volume of the ensemble. To calculate $V$, we must first sum the rate equation over every individual lattice site - 
\begin{equation}
\sum_{i}\dot{N}_{e,i} = -(\eloss+\decayrate)\sum_{i}N_{e,i}-\sum_{i}\frac{K_{ee}}{V_{\text{site}}}N_{e,i}^2,
\end{equation}

where is the volume of an individual lattice site, $V_{\text{site}}$, is assumed to be constant. Additionally assuming that the atoms are approximately distributed normally and the width of the distribution across sites, $\sigma$, does not change significantly with time, we can re-write the above equation as

\begin{equation}
 \dot{\Ne} = -(\eloss+\decayrate)N_e- K_{ee} \frac{N_e^2}{2\sqrt{\pi}\sigma V_{\text{site}}}. \label{eq:two_body} 
\end{equation}

\begin{figure}
    \centering
    \includegraphics[width=\linewidth]{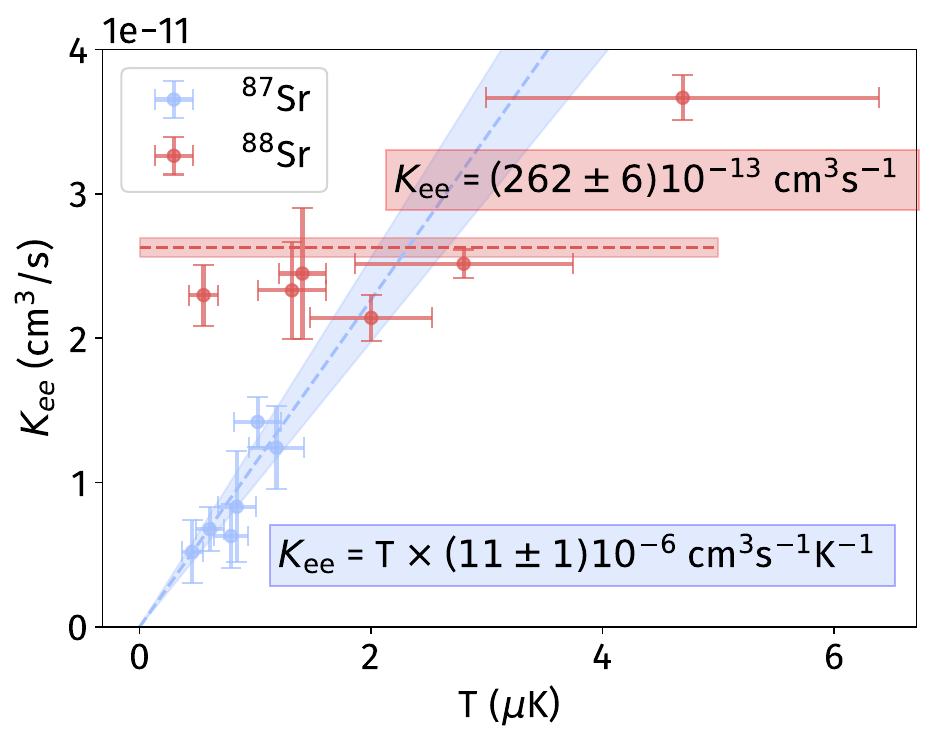}
    \caption{Inferred values of two-body loss $\eloss$ for high-density ensembles of \Sra~and \Srb.}
    \label{fig:two_body}
\end{figure}

Using the fitted values of $\twobodyloss$ and experimentally measured values of $\sigma$ and $V_{\text{site}}$, we extract $K_{ee}$ by comparing the above expression to \cref{eq:rate_equations}, which gives us an effective volume $V = 2\sqrt{\pi}\sigma V_{\text{site}}$. We show these values for the high-density ensemble in \cref{fig:two_body}. For \Sra, we obtain an atomic-temperature-dependent two-body loss rate $K_{ee} =$ \twobodyind. We note that this value is larger than the previously reported value of \twobodyindlit\cite{Bishof11}. For \Srb, we obtain $K_{ee}$ = \twobodybos, which is close, but not entirely consistent with the previously reported value of \twobodyboslit\cite{Traverso09}. A possible reason for the inconsistencies might be that the relatively minor contribution of two-body loss cannot be efficiently decoupled from other losses by the curve fitting routines, due to the high covariance between the various fit parameters. Additionally, the assumptions in \cref{eq:two_body} cease to be valid when individual sites have low atom numbers. Future experiments could leverage greater control of experimental parameters, ideally enabling a triple-zero extrapolation in atom number, trap depth and radial temperature. This would help isolate the effects of various loss channels with low uncertainty.

\begin{table}[]
    \centering
    \begin{tabular}{cccc}
    \hline
    \hline
         \textbf{Rate} & \textbf{This work} & \textbf{Other Expt.} & \textbf{Theory} \\
         
         \hline
         &&&\\
         $\Gamma_{0}$  & 6.0(19) & 6.6(2) \cite{Lu24},  & 6.9(19)\cite{Boyd07} \\
         ($\times10^{-3}$)& & 8.5(2)\cite{Muniz21}, & \\
         (s$^{-1}$)& & 3.0(13)\cite{Dorscher18} & \\
         & & & \\
         $\gamma$ \Sra~& \ramanaunitless &  \dorraman\cite{Dorscher18} & \dorramantheo\cite{Dorscher18}\\
         ($\times10^{-5}$) & & &\\
         (E$_{rec}$s)$^{-1}$ & & & \\
         & & & \\
         $\gamma$ \Srb~& \ramanbunitless & &\\
         ($\times10^{-5}$) & & &\\
         (E$_{rec}$s)$^{-1}$ & & & \\
         & & & \\
         K$_{ee}$ \Sra& \twobodyindunitless & \twobodyindlitunitless \cite{Bishof11}& \\
         ($\times10^{-6}$) & & &\\
         ($ $cm$^{3}$s$^{-1}$K$^{-1}$) & & &\\
         & & & \\
         K$_{ee}$ \Srb& \twobodybosunitless & \twobodyboslitunitless\cite{Traverso09} & \\
         ($\times10^{-13}$) &  &  &\\
         ($ $cm$^{3}$s$^{-1}$) & & &\\
         &&&\\
         \hline
         \hline
         
    \end{tabular}
    \caption{Summary of experimentally measured rates and comparison with existing theory.}
    \label{tab:params}
\end{table}
\section{\label{app:low_density}Low-Density Data}
\begin{figure}
    \centering
    \includegraphics[width=\linewidth]{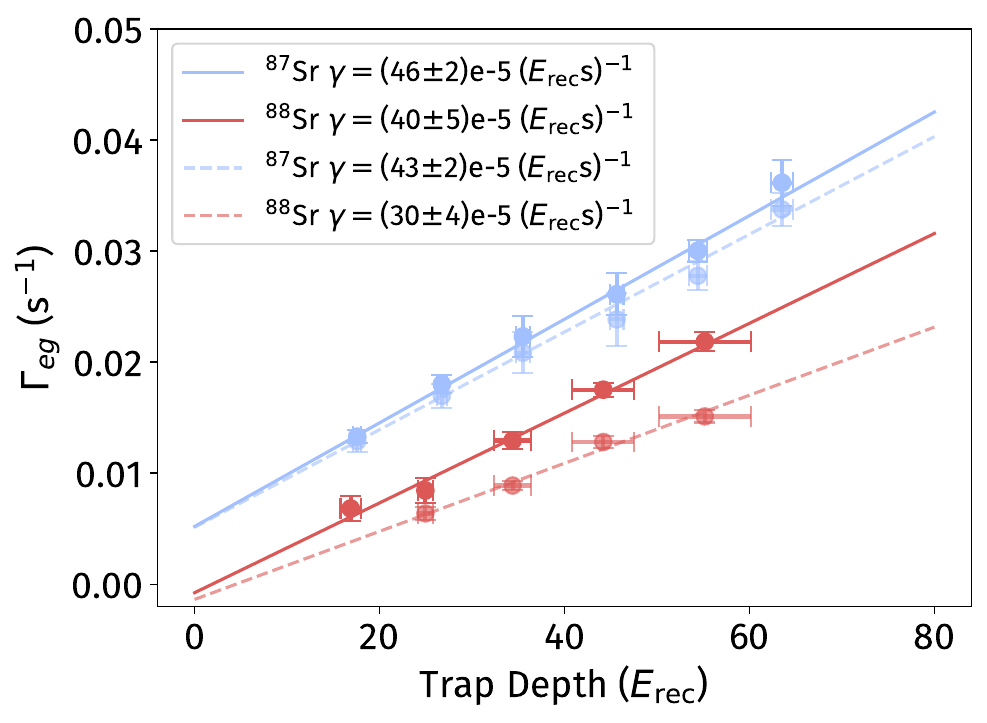}
    \caption{Comparison of extracted decay rates between high (solid) and low (dashed) density ensembles of \Sra~and \Srb~as a function of trap depth.}
    \label{fig:comparison_without_scaling}
\end{figure}

In \cref{fig:comparison_without_scaling}, we observe that while the data from high and low density ensembles of \Sra~are largely consistent with each other, there is a significant difference between the fit parameters for the two ensembles in \Srb. A possible reason for this discrepancy might lie in the calculation of the effective trap depths which depends on the radial temperature of the ensembles. While our analysis assumes that this temperature is the same across both densities, \cref{fig:loss_g} seems to suggest a systematic shift in the ground state loss between high and low densities which could be a result of higher temperatures for the low-density ensembles. This, in turn, would give us a smaller effective trap depth for the low-density data in \Srb, thereby suggesting a larger increase in Raman scattering rate with trap depth. While a thorough analysis of this effect is more suitable for future work, we do observe that if we let the estimated radial temperature of the low-density data to be increased by a factor in the decay model, the decay rates of the two ensembles are more consistent with each other. 

\section{\label{app:bbrrate}Black Body Scattering Rate}
Black body radiation(BBR) can scatter atoms out of the \tripletPzero state primarily through two channels, the E1 transition, \tripletPzero $\rightarrow$ \tripletDone, and the M1 transition, \tripletPzero $\rightarrow$ \tripletPone. 
The BBR scattering rate for each channel be calculated through (Gaussian electromagnetic expressions) 
\begin{equation}
    R_{E1} = \frac{2 \pi}{3\hbar^{2}}\rho(\nu_{E1}, T_{BBR})Z^{2}_{E1},
\end{equation}
\begin{equation}
    R_{M1} = \frac{2 \pi}{3\hbar^{2}}\rho(\nu_{M1}, T_{BBR})Z^{2}_{M1},
\end{equation}
where $\nu_{E1}$ is the \tripletPzero $\rightarrow$ \tripletDone transition frequency, $\nu_{M1}$ is the \tripletPzero $\rightarrow$ \tripletPone transition frequency, $Z_{E1}$ is the reduced matrix element $\braket{^{3}P_{0}||\textbf{D}||^{3}D_{1}}$, $Z_{M1}$ is the reduced matrix element $\braket{^{3}P_{0}||\boldsymbol{\mu}||^{3}P_{1}}$, and $\rho(\nu, T_{BBR})$ is the BBR spectral energy density at frequency $\nu$ and temperature T$_{BBR}$, given by 
\begin{equation}
    \rho(\nu, T_{BBR})= \frac{8 \pi h \nu^{3}}{c^{3}}\frac{1}{e^{h\nu/k_BT}-1}.
\end{equation}
While the reduced matrix element for the electric dipole transition is a factor of $\approx$1/$\alpha$ larger than that of the magnetic dipole transition, the BBR spectral energy density at 300 K is much greater at $\nu_{M1}$ than $\nu_{E1}$. 

Considering both of these channels when calculating the BBR-induced scattering rate out of \tripletPzero, we arrive at a value of \bbrratefull. This value differs from the value reported in \dors et al.\cite{Dorscher18}, as they did not consider the extra contribution from the M1 channel, which amounts to a $\approx$1.7 $\sigma$ correction from their reported value of 2.23(14)$\times$10$^{-3}$ s$^{-1}$.

\clearpage

\newpage
\nocite{apsrev42Control} 

\bibliography{references}


\end{document}